\documentclass[12pt]{article}
\usepackage{amsmath}
\usepackage{bm}

\begin{document}
\title{Exact continuity equation in a space with minimal length}

\author{
H. P. Laba$^1$,
V. M. Tkachuk$^2$\footnote{voltkachuk@gmail.com}\\
$^1$Department of Applied Physics and Nanomaterials Science, \\
Lviv Polytechnic National University,\\
5 Ustiyanovych St., 79013 Lviv, Ukraine,\\
$^2$Department for Theoretical Physics,\\
Ivan Franko National University of Lviv,\\
12, Drahomanov St., Lviv, 79005, Ukraine.}

\maketitle

\begin{abstract}
We derive continuity equation and exact expression for flow probability density
in a space with  arbitrary deformed algebra leading to minimal length. In coordinate representation the flow probability density
is presented as infinite series over parameter of deformation which in momentum representation can be casted into exact closed form determined by deformed kinetic energy.
The flow probability density is calculated explicitly for plane wave and for superposition of two plane waves.


\end{abstract}

\section{Introduction}

The first paper on quantized space-time described by deformed Heisenberg algebra was published in 1947 by Snyder \cite{Sny47}.
For a long time the were  only few papers on this subject. The interest to deformed Heisenberg algebras was renewed after
investigations in String Theory and Quantum Gravity
(see, for example, \cite{Gro88,Mag93,Wit96,Hos12,Hos13}). These studies suggest
existence of nonzero minimal uncertainty in position called minimal length which is of the Plank
scale $l_P=\sqrt{\hbar G/ c^3} \approx 1.6\times 10^{-35} {\rm m} $.
Space with minimal length can be described in the frame of deformed commutation relation for position and momentum (deformed Heisenberg algebra).
One of the simplest deformed Heisenberg algebra leading to the minimal length in one-dimensional case reads
\begin{eqnarray}\label{DefXP}
[\hat X,\hat P]=i\hbar(1+\beta \hat P^2),
\end{eqnarray}
where $\beta$ is parameter of deformation \cite{Kem95,Kem96}.
Applying Heisenberg uncertainty relation to (\ref{DefXP}) one gets the so-called generalized uncertainty principle (GUP). From the principle
follows the following inequality
\begin{eqnarray}
\langle\Delta X^2\rangle\ge \hbar\sqrt\beta,
\end{eqnarray}
where $\langle\Delta X\rangle=\hat X-\langle X\rangle$.
So, in this case we have nonzero minimal uncertainty in position or minimal length which is $\hbar\sqrt\beta$.

We consider more general deformed algebra of the form
\begin{eqnarray}\label{GenXP}
[\hat X,\hat P]=i\hbar F(\sqrt{\beta} \hat P),
\end{eqnarray}
where $F$ is even function. For $\beta=0$ this function reads $F(0)=1$ and we obtain non-deformed algebra.

Algebra (\ref{GenXP}) admits the following representation
\begin{eqnarray}
\hat X=\hat x, \ \ \hat P={1\over\sqrt{\beta}}f(\sqrt{\beta} \hat p),
\end{eqnarray}
where $f$ is odd function,  $\hat x$, and $\hat p$ satisfy non-deformed commutation relation
\begin{eqnarray}
[\hat x,\hat p]=i\hbar.
\end{eqnarray}
Depending on convenience, for small operators $\hat x$, $\hat p$ we can use coordinate representation $\hat x=x$, $\hat p=-i\hbar {\partial \over \partial x}$,
or momentum one  $\hat x=i\hbar {\partial\over \partial p}$, $\hat p=p$.
Domain of changing of momentum is finite $-b\le p\le b$. Note that just this fact
leads to minimal uncertainty in position.
For $\beta\to 0$, we have $f=\sqrt\beta\hat p$ and thus
$\hat P=\hat p$.

It is easy to find that
\begin{eqnarray}
F(\sqrt{\beta}\hat P)=f'(\sqrt{\beta}\hat p)=f'(f^{-1}(\sqrt{\beta}\hat P)),
\end{eqnarray}
here $f^{-1}$ is inverse function to $f$, $f(f^{-1}(x))=x$.
Properties of generalized algebra (\ref{GenXP}) and the minimal length
were studied in details in \cite{Mas12,Now14}.
Note also the paper \cite{Fry03} where generalized three-dimensional deformed algebras were considered.

Choosing different functions of deformation $F$ or respectively different
$f$ we obtain different deformed algebras. For instance, algebra (\ref{DefXP}) has the following representation
\begin{eqnarray}\label{Repres}
\hat P={1\over\sqrt\beta}\tan(\sqrt\beta \hat p), \ \ \hat X=\hat x.
\end{eqnarray}
 Note that according to
(\ref{Repres}) in momentum representation we have
$\pi/2\sqrt \beta \ge p\geq -\pi/2\sqrt \beta $.

Deformation of commutation relations causes additional difficulties in solving many quantum problems. Only for few quantum systems the energy spectra were found exactly in the frame of deformed algebras. They are one-dimensional harmonic oscillator in the space with minimal length \cite{Kem95}  as well as in the space with minimal length and minimal momentum \cite{Que03,Que04}, D-dimensional isotropic harmonic oscillator \cite{Cha02,Dad03},
three-dimensional Dirac
oscillator \cite{Que05}, (2 + 1)-dimensional Dirac oscillator in a homogeneous magnetic field \cite{Men15}, (1 + 1)-dimensional Dirac oscillator within Lorentz-covariant deformed algebra
\cite{Que06}, one-dimensional Coulomb problem \cite{Fit06,Sam16},
the inverse square potential in a space
with  minimal length \cite{Bou07,Bou08}, the (2 + 1)-dimensional Dirac equation in a constant
magnetic field in the presence of  minimal length \cite{Men13,Ped15},
delta potential as well as double delta potential \cite{Sam16Delta},
two-body problems in  space with minimal length \cite{Sam17}.
%%%%%%%%%%%%%%%%%%%%%%%%%
It is worth noting that exact solutions for spectra of physical systems were found in the frame of other types of deformed algebras leading to space quantization. For instance, in the frame of noncommutative algebra of canonical type the energy levels of free particle and harmonic oscillator   \cite{Djemai2004,Kijanka2004,Smailagic2008,Giri2008},  oscillator in a magnetic field \cite{Ben2009,Nath2017}, system of two particles with harmonic oscillator interaction \cite{Gna17},  a particle in a uniform field  \cite{Gna16}  were obtained exactly.

%%%%%%%%%%%%%%%%%%%%%%%%%%%%
Without any doubt continuity equation plays important role in quantum mechanics.
In \cite{Ste07} (see also \cite{Ste08}) 3D deformed algebra with two parameters of deformation $\beta$
and $\beta'$ was considered and in the frame of the algebra the continuity equation, the expression for flow probability density
were obtained in linear approximation over the parameters of deformation. Latter continuity equation for special case $2\beta=\beta'$ (in linear approximation over  parameters of deformation this case corresponds to uniform space) was obtained in \cite{Das09}. These results were used for studies of tunneling effect in 1D deformed space for different barriers. Namely, tunneling in deformed space thought potential step and potential barrier was examined in \cite{Das09}. Tunneling thought delta potential was considered in \cite{Gus18}.
 The authors of paper \cite{Ber16} proposed the approximate method to take into account all orders in parameter of deformation.  The method is based on transformation of higher order differential Schr\"odinger equation in deformed space to second order one.

The aim of this paper is to derive exact continuity equation and exact explicit expression for flow probability density in one-dimensional space with minimal length in general case of deformed algebra.
The paper is organized as follows. In the Section 2 continuity equation and expression for flow probability density are obtained exactly in coordinate representation. Explicit expression for flow probability density in momentum representation is derived in the Section 3. Conclusions are presented in the Section 3.

\section{Schr\"odinger equation and continuity equation}
The time dependent Schr\"odinger equation reads
\begin{eqnarray}
i\hbar{\partial\over\partial t}\psi=H\psi,
\end{eqnarray}
where Hamiltonian in deformed case is supposed to have the same form as is non-deformed one
\begin{eqnarray}
H={\hat P^2\over 2m}+U(\hat X)=T(\hat p,\beta)+U(\hat x).
\end{eqnarray}
Momentum operator and position one satisfy deformed commutation relation (\ref{GenXP}) and can be represented as (\ref{Repres}).

In order to obtain continuity equation we consider coordinate representation
for small operators $\hat x=x$ and $\hat p=-i\hbar {\partial\over \partial x}$.
In this case Hamiltonian reads
\begin{eqnarray}
H=T(\hat p,\beta)+U( x)=T(i\hbar {\partial\over \partial x},\beta)+U( x),
\end{eqnarray}
where operator of kinetic energy is related with deformation function as
\begin{eqnarray}
T(\hat p,\beta)={f^2(\sqrt\beta\hat p)\over 2m\beta}.
\end{eqnarray}
The operator of kinetic energy is even function of $\hat p$  and can be represented as series in $\beta$
\begin{eqnarray}\label{Tser}
T(\hat p,\beta)={1\over 2m}\sum_{n=1}^{\infty}a_n \beta^{n-1}\hat p^{2n},
\end{eqnarray}
where
$a_n={1\over (2n)!}{\partial^{2n} f^2(x)\over \partial x^{2n}}$
at $x=0$.

Let us consider changing in time of the probability density
\begin{eqnarray}\nonumber
{\partial |\psi|^2\over \partial t}={1\over i\hbar}(\psi^*T(\hat p,\beta)\psi-
\psi T(\hat p,\beta)\psi^*)=\\
={1\over 2m i\hbar}\sum_{n=1}^{\infty}\left(a_n \beta^{n-1}
(\psi^*\hat p^{2n}\psi-\psi\hat p^{2n}\psi^*)\right). \label{dtpsi2}
\end{eqnarray}
We rewrite each therm in the right-hand side of this equation as follows.
For the first therm ($n=1$) we have
\begin{eqnarray}
\psi^*\hat p^{2}\psi-\psi\hat p^{2}\psi^*=
\hat p(\psi^*\hat p\psi-\psi\hat p\psi^*).
\end{eqnarray}
The second therm ($n=2$) is rewritten as
\begin{eqnarray}\nonumber
\psi^*\hat p^{4}\psi-\psi\hat p^{4}\psi^*=
\hat p(\psi^*\hat p^3\psi-\psi\hat p^3\psi^*)-
((\hat p\psi^*)\hat p^3\psi-(\hat p\psi)\hat p^3\psi^*))=\\ \nonumber
=\hat p(\psi^*\hat p^3\psi-\psi\hat p^3\psi^*)-
\hat p((\hat p\psi^*)\hat p^2\psi-(\hat p\psi)\hat p^2\psi^*))=\\
=\hat p[\psi^*\hat p^3\psi-\psi\hat p^3\psi^*-
((\hat p\psi^*)\hat p^2\psi-(\hat p\psi)\hat p^2\psi^*))].
\end{eqnarray}

Similarly for the $n$-th therm we obtain
\begin{eqnarray} \nonumber
\psi^*\hat p^{2n}\psi-\psi\hat p^{2n}\psi^*=\\ \nonumber
=\hat p(\psi^*\hat p^{2n-1}\psi-\psi\hat p^{2n-1}\psi^*)-
((\hat p\psi^*)\hat p^{2n-1}\psi-(\hat p\psi)\hat p^{2n-1}\psi^*))=\\ \nonumber
=\hat p(\psi^*\hat p^{2n-1}\psi-\psi\hat p^{2n-1}\psi^*)-
\hat p((\hat p\psi^*)\hat p^{2n-2}\psi-(\hat p\psi)\hat p^{2n-2}\psi^*))+\\ \nonumber
+((\hat p^2\psi^*)\hat p^{2n-2}\psi-(\hat p^2\psi)\hat p^{2n-2}\psi^*))=\\  \nonumber
=\hat p[\psi^*\hat p^{2n-1}\psi-\psi\hat p^{2n-1}\psi^*-
((\hat p\psi^*)\hat p^{2n-2}\psi-(\hat p\psi)\hat p^{2n-2}\psi^*))+\\  \nonumber
+((\hat p^2\psi^*)\hat p^{2n-3}\psi-(\hat p^2\psi)\hat p^{2n-3}\psi^*))]-\\
-((\hat p^3\psi^*)\hat p^{2n-3}\psi-(\hat p^3\psi)\hat p^{2n-3}\psi^*)).
\end{eqnarray}
Continuing we have
\begin{eqnarray}
\psi^*\hat p^{2n}\psi-\psi\hat p^{2n}\psi^*=\hat p\sum_{k=1}^n(-1)^{k-1}
[(\hat p^{k-1}\psi^*)\hat p^{2n-k}\psi-(\hat p^{k-1}\psi)\hat p^{2n-k}\psi^*].
\end{eqnarray}
Using this result we rewrite equation (\ref{dtpsi2}) in the form of continuity equation
\begin{eqnarray} \label{ContEqDef}
{\partial \rho\over \partial t}+{\partial j\over \partial x}=0,
\end{eqnarray}
where $\rho=|\psi|^2$ denotes the  probability density, $j$ is flow probability density which reads
\begin{eqnarray}\label{Genj}
j={1\over 2m}\sum_{n=1}^{\infty}a_n \beta^{n-1}\sum_{k=1}^n(-1)^{k-1}
[(\hat p^{k-1}\psi^*)\hat p^{2n-k}\psi-(\hat p^{k-1}\psi)\hat p^{2n-k}\psi^*].
\end{eqnarray}
Thus we obtain exact expression for the flow probability density in general case of  deformed algebra.
Note that $a_1=1$ and for $\beta=0$ the result for flow probability density reproduces the well known result in  the space with ordinary commutation relation for coordinate and momentum.

\section{Explicit expression for flow probability density in momentum representation}

First of all let us calculate flow probability density for plane wave
$\psi=Ae^{-iT(p,\beta) t/\hbar+ipx/\hbar}$, here $p$  is value of momentum of free particle.
In this case for flow probability density we have
\begin{eqnarray}
j={1\over 2m}|A|^2\sum_{n=1}^{\infty}a_n \beta^{n-1}2np^{2n-1}=
{1\over 2m}|A|^2{\partial\over \partial p}\sum_{n=1}^{\infty}a_n \beta^{n-1}p^{2n}.
\end{eqnarray}
Comparing this result with expression for kinetic energy (\ref{Tser}), we find that
the flow probability density for plane wave can be written in the form
\begin{eqnarray}\label{PlWavej}
j=|A|^2 {\partial T(p,\beta)\over \partial p}.
\end{eqnarray}
Note that according to Hamilton's equations for free particle with zero potential energy we have
\begin{eqnarray}\label{vp}
{\partial T(p,\beta)\over \partial p}={dx\over dt}=v,
\end{eqnarray}
$v$ is velocity of particle. Thus the flow probability density reads
\begin{eqnarray}\label{jv}
j=|A|^2v.
\end{eqnarray}
This expression is similar to expression for the flow probability density in non-deformed case. The only one difference is that relation of velosity of the particle with momentum is deformed according to (\ref{vp}).

Now we consider  wave function which is superposition of plane waves
\begin{eqnarray}
\psi(x)=\int_{-b}^{b} dp c(p) e^{ipx/\hbar},
\end{eqnarray}
where we take into account that domain of changing  of momentum is finite.
Substituting it into (\ref{Genj}), we find
\begin{eqnarray} \nonumber
j=\int_{-b}^{b}dp\int_{-b}^{b} dq c(p) c^*(q)e^{i(p-q)x/\hbar}\times\\ \nonumber
\times {1\over 2m}\sum_{n=1}^{\infty}a_n \beta^{n-1}
\sum_{k=1}^n(-1)^{k-1}[(-q)^{k-1}p^{2n-k}-p^{k-1}(-q)^{2n-k}]=\\ \nonumber
=\int_{-b}^{b}dp\int_{-b}^{b} dq c(p) c^*(q)e^{i(p-q)x/\hbar}\times\\
\times {1\over 2m}\sum_{n=1}^{\infty}a_n \beta^{n-1}
\sum_{k=1}^n[q^{k-1}p^{2n-k}+p^{k-1}q^{2n-k}].
\end{eqnarray}
It is interesting that this result can be written in closed form. For this we note
that the sum over $k$ can be done explicitly using geometric progression
\begin{eqnarray}
\sum_{k=1}^n[q^{k-1}p^{2n-k}+p^{k-1}q^{2n-k}]={p^{2n}-q^{2n}\over p-q}.
\end{eqnarray}
Then we have
\begin{eqnarray}
{1\over 2m}\sum_{n=1}^{\infty}a_n \beta^{n-1}{p^{2n}-q^{2n}\over p-q}=
{T(p,\beta)-T(q,\beta)\over p-q}.
\end{eqnarray}
Finally, the closed expression for flow probability density in the space with
deformed algebra reads
\begin{eqnarray}\label{Exactj}
j=\int_{-b}^{b}dp\int_{-b}^{b} dq c(p) c^*(q)
{T(p,\beta)-T(q,\beta)\over p-q}e^{i(p-q)x/\hbar}.
\end{eqnarray}

Let us consider non-deformed case, namely $\beta=0$. In this case $T(p,0)=p^2/2m$ and
the domain of $p$ is real line.
So, we find
\begin{eqnarray}
j={1\over 2m}\int_{-\infty}^{\infty}dp\int_{-\infty}^{\infty} dq c(p) c^*(q)
(p+q)e^{i(p-q)x/\hbar},
\end{eqnarray}
that reproduces the flow probability density in non-deformed case.

Now let us calculate flow probability density in deformed case according to (\ref{Exactj}) for
a few examples of wave functions.
As a first example we consider plane wave, which is solution of time-dependent Schr\"odinger equation for free particle.   In momentum representation it reads
\begin{eqnarray}
c(p)=Ae^{-iT(p,\beta) t/\hbar}\delta(p-p_0),
\end{eqnarray}
where $p_0$ is momentum of plane wave. Substituting it into (\ref{Exactj})
we reproduce result (\ref{PlWavej})
for flow probability density of plane wave with momentum $p_0$.

As a second example let us consider superposition of two plane waves with
momenta $p_1$ and $p_2$. In this case wave function in momentum representation satisfying
time-dependent Schr\"odinger equation
reads
\begin{eqnarray}
c(p)=Ae^{-iT(p_1,\beta) t/\hbar}\delta(p-p_1)+Be^{-iT(p_2,\beta) t/\hbar}\delta(p-p_2),
\end{eqnarray}
where $A=|A|e^{i\phi_1}$, $B=|B|e^{i\phi_2}$.
Then for flow probability density we find
\begin{eqnarray}\nonumber
 j=j_{p_1}+j_{p_2}+ {\Delta T\over \Delta p}
 \left(AB^*e^{-i\Delta T t/\hbar}e^{i\Delta p x/\hbar}
 +A^*Be^{i\Delta T t/\hbar}e^{-i\Delta p x/\hbar}\right)= \\ \label{TwoPlj}
 =j_{p_1}+j_{p_2}+ {\Delta T\over \Delta p}
2|A||B|\cos(\Delta p x/\hbar-\Delta T t/\hbar+\Delta\phi),
\end{eqnarray}
here we introduce the notations $\Delta T=T(p_1,\beta)-T(p_2,\beta)$,
$\Delta p=p_1-p_2$, $\Delta\phi=\phi_1-\phi_2$.
Note that in the case $p_2\to p_1$ we recover the result for
flow probability density of plane wave with momentum $p_1$ and
amplitude $A+B$ as it must be. Similarly as in non-deformed case, for $p_2=-p_1$ we find that
in the frame of deformed algebra
$j=j_{p_1}+j_{-p_1}$ and interference term is absent.

\section{Conclusions}
In this paper we have studied time-dependent Schr\"odinger equation in a space with minimal length with arbitrary deformed algebra.
We have derived exact continuity equation (\ref{ContEqDef}) and exact expression for flow probability density in coordinate representation in general case of deformed algebra leading to minimal length (\ref{Genj}).  The flow probability density has been written in the form of series in parameter of deformation $\beta$. It is interesting that
in momentum representation the flow probability density can be written in explicit closed form
(\ref{Exactj}) and it is determined by the kinetic energy of particle in deformed space.

 The obtained results allows exact calculation of the flow probability density in spaces with deformed algebras for a given wave function. As an example we have found explicitly the flow probability density for one plane wave (\ref{PlWavej}) and for superposition of two plane waves (\ref{TwoPlj}). The flow probability density for one plane wave in space with deformed algebra is proportional to velosity of particle (\ref{jv}) as in nondeformed case. The difference is that relation of velosity of the particle with momentum is deformed according to (\ref{vp}).
In the case of superposition of two plane waves we have shown that when plane waves have the same wave vectors with opposite direction, the total flow probability density is equal to sum of flows corresponding to the first and the second waves. The total flow probability density for arbitrary two plane  waves  contains additional interference therm.
Finally, it is worth noting  that  obtained results give the way for exact
 studies of the tunneling effect in space with arbitrary deformed algebras leading to the minimal length.

\section*{Acknowledgments}
 This work was supported by the Project $\Phi\Phi$-83$\Phi$
(No. 0119U002203) from the Ministry of Education and
Science of Ukraine. The authors thank Dr. Kh. Gnatenko for useful comments.

\end{document}